\documentclass{JHEP3}
\pdfoutput=1

\usepackage{graphicx}
\usepackage{amsmath,amsfonts,bbm,times,slashed}

\DeclareMathOperator{\re}{Re}
\DeclareMathOperator{\tr}{Tr}
\DeclareMathOperator{\SU}{SU}
\DeclareMathOperator{\Gl}{Gl}
\DeclareMathOperator{\U}{U}
\DeclareMathOperator{\1}{\mathbbm{1}}
\DeclareMathOperator{\diag}{\textrm{diag}}
\DeclareMathOperator{\sdet}{Sdet}
\DeclareMathOperator{\str}{Str}

\newcommand{\R}{\mathbbm{R}}
\newcommand{\bDelta}{{\bar\Delta}}
\newcommand{\Ds}{\slashed{D}}
\newcommand{\braket}[1]{\left\langle #1 \right\rangle}

\renewcommand\epsilon\varepsilon
\renewcommand\phi\varphi
\renewcommand{\O}{\mathcal{O}}
\renewcommand{\L}{{\cal L}}

\title{Partially quenched chiral perturbation theory in the epsilon
  regime at next-to-leading order}
\author{Christoph Lehner and Tilo Wettig\\
  Institute for Theoretical Physics, University of Regensburg, 93040
  Regensburg, Germany\\
  Email: \email{christoph.lehner@physik.uni-regensburg.de},
  \email{tilo.wettig@physik.uni-regensburg.de}}

\abstract{We calculate the partition function of partially quenched
  chiral perturbation theory in the epsilon regime at next-to-leading
  order using the supersymmetry method in the formulation without a
  singlet particle.  We include a nonzero imaginary chemical potential and
  show that the finite-volume corrections to the low-energy constants
  $\Sigma$ and $F$ for the partially quenched partition function, and
  hence for spectral correlation functions of the Dirac operator, are
  the same as for the unquenched partition function.  We briefly 
  comment on how to minimize these corrections in lattice simulations
  of QCD.  As a side result, we show that the zero-momentum integral
  in the formulation without a singlet particle agrees with previous results
  from random matrix theory.}

\keywords{partially quenched chiral perturbation theory, imaginary chemical potential,
  supersymmetry method, finite-volume corrections, low-energy constants}

\preprint{October 13, 2009}

\begin{document}

\section{Introduction}

At low energies, QCD can be described by a chiral effective theory.
If the theory is considered in a finite volume and for small quark
masses, the standard $p$-regime power counting is replaced by the
$\epsilon$-regime power counting introduced by Gasser and Leutwyler
\cite{Gasser:1987ah}.  To leading order in the $\epsilon$-regime, the
partition function is dominated by the contribution of the
zero-momentum modes of the Nambu-Goldstone (NG) bosons
\cite{Gasser:1987ah,Leutwyler:1992yt}.  In this limit the theory
becomes zero-dimensional and is therefore described by chiral random
matrix theory (RMT) \cite{Shuryak:1992pi}, see
\cite{Verbaarschot:2000dy,Akemann:2007rf} for reviews.  The low-energy
constants (LEC) appearing in the chiral effective Lagrangian, which
are of great phenomenological importance, can be determined by fitting
analytical results from RMT to lattice data for the eigenvalue
spectrum of the Dirac operator.  The lowest-order LECs are $\Sigma$
and $F$.  While $\Sigma$ can be determined rather easily, e.g., from
the distribution of the small Dirac eigenvalues, the extraction of $F$
is somewhat more complicated and requires the inclusion of a suitable
chemical potential \cite{Damgaard:2005ys,Akemann:2006ru}.

Since lattice simulations are restricted to a finite volume, it is
important to take into account finite-volume corrections to the RMT
results, which can be obtained by going to next-to-leading order (NLO) in
the $\epsilon$-regime.
Recently, finite-volume corrections to the unquenched partition function of QCD
in the $\epsilon$-regime were obtained in
\cite{Damgaard:2007xg,Akemann:2008vp}.  However, in order to extract
the relevant eigenvalue correlation functions the partially quenched
partition function of QCD is needed.  A relatively simple method to
obtain the partially quenched theory is to introduce $n$ replicated
flavors in the unquenched theory and then to analytically continue in
the discrete number of quark flavors to zero.  This so-called replica
trick was first used in the theory of disordered systems
\cite{Edwards:1975zz}.  It is potentially problematic
since the analytic continuation from an isolated
set of points is not uniquely defined.  Nevertheless, a number of authors
have succeeded to construct proper analytic continuations from
which correct results could be obtained, see, e.g.,
\cite{Kamenev:1999aa,Kanzieper:2002ix,Splittorff:2002eb}.
Several publications in QCD
have used the replica trick for perturbative calculations while
borrowing exact result for the non-perturbative part of the theory
from RMT
\cite{Damgaard:2007ep,Bernardoni:2007hi,Bernardoni:2008ei,Damgaard:2008zs}.

In this publication we choose to use an alternative way to obtain the
partially quenched theory that does not suffer from the potential
problems of the replica trick and can therefore be used to check and
extend previous results.  In addition to the sea quarks, we introduce
fermionic and bosonic valence quarks.  In nuclear physics and
condensed matter physics this method is known as the supersymmetry
method or Efetov method for quenched disorder \cite{Efetov:1983zz}.
In the context of QCD this idea was first used by Morel
\cite{Morel:1987xk}.  The effective low-energy theory of QCD with
$N_f+N_v$ quarks and $N_v$ bosonic quarks was developed by Bernard and
Golterman \cite{Bernard:1993sv} and by Sharpe and Shoresh
\cite{Sharpe:2001fh}.  In this work we use the effective theory
without a singlet particle as
discussed by Sharpe and Shoresh and consider it in a finite volume and
for small quark masses.  In order to access $F$ in addition to
$\Sigma$, we include an imaginary quark chemical potential $\mu$
\cite{Damgaard:2005ys,Akemann:2006ru}.  (A first exploratory lattice
study of this idea was performed in Ref.~\cite{DeGrand:2007tm}.)  We
compute the partition function at next-to-leading order in the
$\epsilon$-regime and thereby obtain finite-volume corrections of
order $1/\sqrt{V}$ to the partially quenched theory that translate
into finite-volume corrections to the LECs $\Sigma$ and $F$.  Our
results agree with previous results for the unquenched partition
function \cite{Gasser:1987ah,Damgaard:2007xg,Akemann:2008vp}.  As a side result we
demonstrate that the parametrization of the NG manifold by Sharpe and
Shoresh leads to the correct universal limit, in analogy to the
results of Refs.~\cite{Osborn:1998qb,Damgaard:1998xy} where a
different parametrization was used.

An important question is to what extent the finite-volume effects
in the determination of a particular quantity, such as $\Sigma$ or $F$,
are universal in the sense that different methods used to determine this quantity
give rise to the same finite-volume effects.  In general the
effects of the finite volume depend on the method, see, e.g., the
finite-volume effects in the determination of $F$ in Ref.~\cite{Fukaya:2007pn}.
In the present paper we show that at next-to-leading order in the $\varepsilon$-expansion
the partially quenched partition function is equal to its infinite-volume
counterpart with $\Sigma$ and $F$ replaced by effective values $\Sigma_\text{eff}$
and $F_\text{eff}$.  Since the knowledge of the analytic form of the
partially quenched partition function suffices to determine all spectral correlation
functions of the Dirac operator $\Ds$ we find that all quantities
that can be expressed in terms of spectral correlation functions of $\Ds$
give rise to the same finite-volume corrections to $\Sigma$ and $F$.

This paper is structured as follows.  In section~\ref{sec:pq} we
review the partially quenched theory and how it can be used to compute
spectral correlation functions.  We also review the corresponding
effective low-energy theory in the formulation of Sharpe and Shoresh,
both at fixed vacuum angle $\theta$ and at fixed topology $\nu$.  In
section~\ref{sec:eps} we compute the finite-volume corrections of
order $1/\sqrt{V}$ to the partially quenched theory, and thus to
$\Sigma$ and $F$.  We also show that the correct universal limit is
obtained from the formulation of Sharpe and Shoresh.  Conclusions are
drawn in section~\ref{sec:concl}.  An appendix is provided to collect
some useful formulas for the massless propagator in dimensional
regularization, including commonly used shape coefficients.

\section{\boldmath QCD with $N_f+N_v$ quarks and $N_v$ bosonic quarks in a
  finite volume}\label{sec:pq}

In this section we consider QCD with $N_f+N_v$ quarks and $N_v$
bosonic quarks (Morel's bosonic spin-$1/2$ ghost fields
\cite{Morel:1987xk}) in a box of volume $V=L_0 L_1 L_2 L_3$ in the
Euclidean formalism.  The temporal extent of the box is given by
$L_0$, and thus the temperature of the system is $T=1/L_0$.
Unless stated otherwise we consider the partially quenched case of
$N_f>0$.

\subsection{The partition function and spectral correlation functions}

We define QCD with $N_f+N_v$ quarks and $N_v$ bosonic quarks
by the partition function
\begin{align} \label{eqn:partqcd}
  Z = \int d[A] \: e^{-S_\text{YM}}
  \Biggl[\prod_{f=1}^{N_f} \det(\Ds+m_f)\Biggr]
  \Biggl[\prod_{i=1}^{N_v}\frac{\det(\Ds+m_{vi})}
  {\det(\Ds+m'_{vi})}\Biggr] \,,
\end{align}
where the integral is over all gauge fields $A$, $S_\text{YM}$ is the
Yang-Mills action, $\Ds$ is the Dirac operator, $m_1,\ldots,m_{N_f}$
are the masses of the sea quarks, $m_{v1},\ldots,m_{vN_v}$ are the
masses of the fermionic valence quarks, and $m'_{v1},\ldots,m'_{vN_v}$
are the masses of the bosonic valence quarks.  By setting the mass
$m_{vi}$ of a valence quark equal to the mass $m'_{vi}$ of the
corresponding bosonic quark, the ratio of determinants of this pair
cancels and the flavor $i$ is quenched.

Next we rewrite the determinants in terms of fermionic quark fields
$\psi$ and bosonic quark fields $\phi$ using
\begin{align}\label{eqn:ferdet}
  \det(\Ds+m) = \int d[\bar\psi\psi] \: e^{-\int d^4x\, \bar\psi
    (\Ds+m)\psi }
\end{align}
and
\begin{align}\label{eqn:bosdet}
  \frac1{\det(\Ds+m)} = \int d[\bar\phi\phi] \: e^{-\int d^4x\,
    \bar\phi (\Ds+m)\phi }\,,
\end{align}
where $\psi$ and $\bar\psi$ are independent Grassmann variables with
Berezin integral $\int d[\bar\psi\psi]$, and $\phi$ and $\bar\phi$ are
commuting complex fields related by complex conjugation, $\bar\phi =
\phi^\dagger$.  The integrals in the exponents are over space-time.
Note that the right-hand side of Eq.~\eqref{eqn:bosdet} only converges if all
eigenvalues of $\Ds+m$ have a positive real part.  Since $\Ds$ is
anti-Hermitian this condition is satisfied as long as $\re m > 0$.
Thus 
\begin{align}\label{eqn:partqcdf}
  Z = \int d[A] \, d[\bar\Psi\Psi]\: e^{- S_\text{YM} -\int d^4x\,
    \bar\Psi (\Ds + M)\Psi}
\end{align}
with mass matrix
$M=\diag(m_1,\ldots,m_{N_f},m_{v1},\ldots,m_{vN_v},m'_{v1},\ldots,m'_{vN_v})$
and fields
\begin{align}
  \bar\Psi =
  \begin{pmatrix}
    \bar\psi &
    \bar\phi
  \end{pmatrix},\qquad
  \Psi =
  \begin{pmatrix}
    \psi \\
    \phi
  \end{pmatrix}.
\end{align}
At nonzero temperature we have to choose anti-periodic boundary
conditions in the temporal direction for the fermionic quarks.
A pair of fermionic and bosonic quarks at equal mass has
to cancel in Eq.~\eqref{eqn:partqcd}, and therefore
we have to choose anti-periodic boundary conditions in the temporal
direction also for the bosonic quarks 
(in the same way as Faddeev-Popov ghosts acquire
periodic boundary conditions at nonzero temperature
\cite{Bernard:1974bq}).  This will amount to periodic boundary
conditions for pseudo-NG fermions composed of quarks and bosonic
anti-quarks (or of anti-quarks and bosonic quarks).  

The vacuum expectation value of an operator $\O$ is given by
\begin{align}
  \braket{ {\cal O} } = \frac1{Z} \int d[A] \, d[\bar\Psi\Psi]\: \O\:
  e^{-S_\text{YM}-\int d^4x\, \bar\Psi(\Ds+M)\Psi}\,. 
\end{align}
For example, choosing $N_v=1$, the presence of a bosonic quark can be
used to obtain the spectral density (or one-point function) of the
Dirac operator $\Ds$,
\begin{align}
  \rho(\lambda) = \braket{\tr \delta(\Ds-i\lambda)}
  = \lim_{\epsilon\to0}\frac1{\pi} 
  \re \braket{\tr (\Ds-i\lambda+\varepsilon)^{-1}} \,,
\end{align}
by using
\begin{align}
  \braket{\tr (\Ds+m)^{-1}} = \left.\frac{\partial}{\partial m_v} 
    \log Z(m_1,\ldots,m_{N_f};m_v,m'_v)\right|_{m_v=m_v'=m}\,.
\end{align}
Analogously, higher-order spectral correlation functions can be
obtained using $N_v=k$, where $k$ is the desired order.  From these
$k$-point functions we can also compute individual eigenvalue
distributions \cite{Akemann:2003tv}.

\subsection[The effective low-energy theory at fixed vacuum angle
$\theta$]{\boldmath The effective low-energy theory at fixed vacuum
  angle $\theta$}

In this section we briefly discuss how to determine the relevant
low-energy degrees of freedom for QCD with $N_f+N_v$ quarks and $N_v$
bosonic quarks.  For details we refer to Ref.~\cite{Sharpe:2001fh}.
The general procedure is as follows.  We first determine the
non-anomalous symmetries of the Lagrangian that act non-trivially on
the vacuum.  Then we restrict the remaining symmetry generators to a
subset that is sufficient to generate all Ward identities associated
with the flavor symmetries.  This subset of symmetry generators then
determines the relevant NG manifold of the effective low-energy
theory.  The Lagrangian of the quark sector is given by
\begin{align}
  \L_Q = \bar\Psi (\Ds + M) \Psi\,,
\end{align}
which in the massless case ($M=0$) has vector and axial symmetries.
The vacuum of this theory is invariant under the vector symmetry.  The
axial symmetry, however, acts non-trivially on the vacuum.\footnote{
  For the detailed arguments concerning the symmetry breaking pattern
  of QCD with $N_f+N_v$ quarks and $N_v$ bosonic quarks we again refer
  to Ref.~\cite{Sharpe:2001fh}.}  The axial symmetry is defined by a
supermanifold \cite{Zirnbauer:1996zz}
with base
\begin{align}
 \Gl(N_f+N_v) \otimes [\Gl(N_v) / \U(N_v)]\,,
\end{align}
where $\Gl$ is the general linear group and $\U$ is its unitary
subgroup.  The factor $\Gl(N_f+N_v)$ acts on the quark sector while
$\Gl(N_v) / \U(N_v)$ acts on the bosonic quark sector
\cite{Osborn:1998qb,Damgaard:1998xy}.  The reason for the smaller
symmetry group of the bosonic quark sector is that $\phi$ and $\bar\phi$ are
related by complex conjugation, while $\psi$ and $\bar\psi$ are
independent in the functional integral.
The measure of the functional integral restricted to
the topological sector $\nu$ transforms under axial transformations
$U_A$ as \cite{Dalmazi:2000bs}
\begin{align} \label{eqn:toptrafo}
  d[\bar\Psi\Psi] \to \sdet^\nu(U_A) \, d[\bar\Psi\Psi]\,,
\end{align}
where $\sdet$ is the superdeterminant \cite{Efetov:1983zz}.  Thus,
for $\nu \ne 0$, only axial transformations with $\sdet(U_A)=1$ leave the measure invariant,
i.e., are non-anomalous.  Let us express an arbitrary axial
transformation $U_A$ by
\begin{align}
  U_A = \exp \left( i G_A \right) = \exp i
  \begin{pmatrix}
    u_A & \bar\kappa^T \\
    \kappa & u'_A
  \end{pmatrix},
\end{align}
where $\kappa$ and $\bar\kappa$ are independent $N_v \times (N_f+N_v)$
matrices with elements in the Grassmann algebra, $u_A$ lives in the group algebra
of $\Gl(N_f+N_v)$,
and $u'_A$ lives in the group algebra of $\Gl(N_v) / \U(N_v)$.  The
restriction $\sdet(U_A)=1$ amounts to the requirement of a vanishing
supertrace \cite{Efetov:1983zz} of $G_A$, i.e., $\str G_A = \tr u_A -
\tr u'_A=0$.  Next we restrict the remaining axial symmetries to the
minimal subset that is necessary to generate all Ward identities of
the full symmetry.  Note that $\Gl(N_f+N_v)$ contains the same
generators as $\U(N_f+N_v)$ with real coordinates replaced by complex
ones.  Since this does not give rise to additional Ward identities it
is sufficient to keep either the real or the imaginary part of each
coordinate.  The choice made in Ref.~\cite{Sharpe:2001fh} is
\begin{align}\label{eqn:defga}
 G_A =
  \begin{pmatrix}
    \pi & \bar \kappa^T \\
    \kappa & i\pi'
  \end{pmatrix}
  + \frac{i \phi}{\sqrt{(N_f+N_v) N_v N_f}}
  \begin{pmatrix}
    N_v \1_{N_f+N_v} & 0 \\
    0 & (N_f+N_v) \1_{N_v}
  \end{pmatrix},
\end{align}
where $\pi=\pi^\dagger$ and $\pi'=\pi'^\dagger$ are traceless
Hermitian matrices of dimension $N_f+N_v$ and $N_v$, respectively,
$\phi \in \R$, and $\1_n$ is the $n$-dimensional identity matrix.
This choice leads to the correct signs of the kinetic terms of the NG
particles in the effective low-energy theory and will also be used in
the rest of this paper.  Note that for $N_f=0$ also the flavor singlet
particle will give rise to long-range correlations
\cite{Sharpe:2001fh} and thus has to be included in the effective
theory.

The transformation properties of the massive theory under axial
transformations as well as the Lorentz group now dictate the form of
the Lagrangian of the effective theory \cite{Dalmazi:2000bs}.  To
leading order in $U(x)$, $\partial_\rho U(x)$, and $M$ we find
\begin{align}
  \L_\text{eff} = \frac{F^2}4 \str\, \bigl[ \partial_\rho
  U(x)^{-1} \partial_\rho U(x) \bigr] - \frac{\Sigma}{2} \str\,\bigl[
  M^\dagger U(x) + U(x)^{-1} M\bigr]\,,
\end{align}
where $F$ and $\Sigma$ are low-energy constants and the NG manifold
$U(x)$ is obtained by promoting the coordinates $\pi$, $\pi'$,
$\kappa$, $\bar\kappa$, and $\phi$ in Eq.~\eqref{eqn:defga} to fields
with
\begin{align}
  U(x) = \exp \bigl( i G_A(x) \bigr)\,.
\end{align}
The theory in a $\theta$-vacuum is then obtained by rotating the sea
quark masses,
\begin{align}
  \L_\text{eff}(\theta) = \frac{F^2}4 \str\, \bigl[ \partial_\rho
  U(x)^{-1} \partial_\rho U(x) \bigr] - \frac{\Sigma}{2} \str\,\bigl[
  M^\dagger e^{-i \bar\theta / N_f} U(x) + U(x)^{-1} e^{i \bar\theta /
    N_f} M\bigr]\,,
\end{align}
where
\begin{align}\label{eqn:defbartheta}
  \bar \theta =
  \theta \begin{pmatrix}
    \1_{N_f} & 0 \\
    0 & 0
  \end{pmatrix}
\end{align}
is an $(N_f+2N_v)$-dimensional matrix that projects onto the sea-quark
sector.  The partition function of the effective
theory at fixed $\theta$ is thus given by
\begin{align} \label{eqn:parttheta} 
  Z_\text{eff}(\theta) = \int d[U] \: e^{-\int d^4x\,
    \L_\text{eff}(\theta)}\,,
\end{align}
where $d[U]$ is the invariant integration measure associated with the
supermanifold \cite{Zirnbauer:1996zz}.  We restrict ourselves to the
effective theory in the rest of this paper and thus drop the subscript
in the following.

\subsection[The effective low-energy theory at fixed topology
$\nu$]{\boldmath The effective low-energy theory at fixed topology
  $\nu$}

The partition function at fixed $\theta$-angle is given by the Fourier
series
\begin{align}
  Z(\theta) = \sum_{\nu = -\infty}^\infty e^{i \theta \nu} Z_\nu \,,
\end{align}
and thus the partition function at fixed topological charge $\nu$ is
obtained by the Fourier transform
\begin{align}
  Z_\nu = \frac1{2\pi}\int_0^{2\pi} d\theta \: e^{-i\theta \nu}
  Z(\theta)\,.
\end{align}
For the partition function defined in Eq.~\eqref{eqn:parttheta} this
means
\begin{align}\label{eqn:Znu}
  Z_\nu = \int d\theta \int d[U]\: \exp\,&\biggl\{-i\theta \nu
  -\int d^4x\, \biggl(\frac{F^2}4  \str\, \bigl[ \partial_\rho
  U(x)^{-1} \partial_\rho U(x) \bigr] \notag\\
  & \qquad\qquad - \frac{\Sigma}{2} \str\,\bigl[ M^\dagger e^{-i
    \bar\theta / N_f} U(x) + U(x)^{-1} e^{i \bar\theta / N_f}
  M\bigr]\biggr) \biggr\}\,.
\end{align}
If we separate the constant mode $U_0$ from $U(x)$ by the
ansatz
\begin{align}
  U(x) = U_0 \exp \bigl( i G_A(x) \bigr)
\end{align}
with $\int d^4x\: G_A(x) = 0$ and $U_0 = \exp ( i G_A^0)$, we can
absorb $\theta$ in $U_0$ by
\begin{align}
  \label{eqn:pitilde}
  \pi_0 \to \tilde\pi_0=\pi_0 - \frac{\theta}{N_f} \begin{pmatrix}
    \1_{N_f} & 0 \\
    0 & 0
  \end{pmatrix},
\end{align}
where $\pi_0$ is the constant mode of the pion fields in the fermionic
quark sector of $G_A^0$.  To avoid confusion with \eqref{eqn:defbartheta}
we mention that the matrix in \eqref{eqn:pitilde} has dimension
$N_f+N_v$.  Note that we absorb the $\theta$-angle only in the sea
sector of the theory.  This yields
\begin{align}
  Z_\nu = \int d[U]\: \sdet^\nu(U_0) \exp\,\biggl\{
  -\int d^4x\,\biggl(&\frac{F^2}4  \str\, \bigl[ \partial_\rho
  U(x)^{-1} \partial_\rho U(x) \bigr] \notag\\
  & - \frac{\Sigma}{2} \str\,\bigl[ M^\dagger U(x) +
  U(x)^{-1}M\bigr]\biggr)\biggr\}\,,
\end{align}
where the integration manifold for the constant mode is changed from
\eqref{eqn:defga} to
\begin{align}\label{eqn:defgaf}
  G_A^0 =
  \begin{pmatrix}
    \tilde\pi_0 & \bar \kappa^T_0 \\
    \kappa_0 & i\pi_0'
  \end{pmatrix}
  + \frac{i \phi_0}{\sqrt{(N_f+N_v) N_v N_f}}
  \begin{pmatrix}
    N_v \1_{N_f+N_v} & 0 \\
    0 & (N_f+N_v) \1_{N_v}
  \end{pmatrix},
\end{align}
in which $\tilde\pi_0$ now generates $\U(N_f+N_v)$ instead of
$\SU(N_f+N_v)$\footnote{
The addition of $\1_{N_f}$ to the generators of $\SU(N_f+N_v)$ suffices to generate $\U(N_f+N_v)$.
The normalization of $\theta$ in Eq.~\eqref{eqn:pitilde} yields the correct integration domain.
} while $\pi_0'$, $\bar\kappa_0$, $\kappa_0$
and $\phi_0$ are defined in the same way
as their counterparts in Eq.~\eqref{eqn:defga}.
Note that this parametrization of the constant mode
is different from the parametrization used previously in the
literature \cite{Osborn:1998qb,Damgaard:1998xy}.  In
section~\ref{sec:universal} we will show that this parametrization
again yields the universal RMT result.

\section{Finite-volume corrections}\label{sec:eps}

\subsection[The $\varepsilon$-expansion in the effective theory with
imaginary chemical potential]{\boldmath The $\varepsilon$-expansion in
  the effective theory with imaginary chemical potential}

For convenience we redefine the NG manifold with a different
normalization of the fields by
\begin{align}\label{eqn:defu}
  U(x) = U_0 \exp\biggl(\frac{i\sqrt 2}{F} \xi(x) \biggr)
\end{align}
with
\begin{align}\label{eqn:defxi}
  \xi(x) =
  \begin{pmatrix}
    \pi(x) & \bar \kappa^T(x) \\
    \kappa(x) & i\pi'(x)
  \end{pmatrix}
  + \frac{i \phi(x)}{\sqrt{(N_f+N_v) N_v N_f}}
  \begin{pmatrix}
    N_v \1_{N_f+N_v} & 0 \\
    0 & (N_f+N_v) \1_{N_v}
  \end{pmatrix}.
\end{align}
The constant mode is separated in $U_0$, and thus $\int d^4x\, \xi(x)
= 0$.  For nonzero imaginary chemical potential the Lagrangian of the
effective theory is given by
\begin{align}
  {\cal L} = \frac{F^2}4 \str\, \bigl[ \nabla_\rho U(x)^{-1}
  \nabla_\rho U(x) \bigr] - \frac{\Sigma}{2} \str\,\bigl[ M^\dagger
  U(x) + U(x)^{-1} M\bigr]
\end{align}
with
\begin{align}\label{eqn:nablachem}
  \nabla_\rho U(x) = \partial_\rho U(x) -i \delta_{\rho 0}[C,U(x)]\,,
\end{align}
where $C=\diag(\mu_1,\ldots,\mu_{N_f},\mu_{v1},\ldots,\mu_{vN_v},
\mu'_{v1},\ldots,\mu'_{vN_v})$ and $i\mu_i$ is the imaginary chemical
potential of quark flavor $i$.  We use the $\varepsilon$-regime power
counting \cite{Gasser:1987ah} defined by
\begin{align}
  V \sim \varepsilon^{-4}\,,\qquad M \sim \varepsilon^4\,,\qquad \mu
  \sim \varepsilon^2\,,\qquad
  \partial_\rho \sim \varepsilon\,,\qquad \xi(x) \sim \varepsilon\,.
\end{align}
Note that the expansion in $\varepsilon^2$ amounts to an expansion in
$1/\sqrt{V}$.  To leading order in $\varepsilon^2$ the Lagrangian is
given by
\begin{align}\label{eqn:lagrlo}
  {\cal L}_0 &= \frac{1}{2} \str\, \bigl[\partial_\rho\xi(x)
  \partial_\rho\xi(x)\bigr] -\frac\Sigma 2\str\,\bigl[ M^\dagger U_0 +
  U_0^{-1} M \bigr] -\frac{F^2}4\str\, [C, U_0^{-1}] [C, U_0]\,.
\end{align}
The next-to-leading order terms in $\varepsilon^2$ are
\begin{align}\label{eqn:lagrnlo}
  {\cal L}_2 &= {\cal L}^M_2 + {\cal L}^C_2 + {\cal L}^N_2
\end{align}
with
\begin{align}
  \label{eq:L2M}
  {\cal L}^M_2 &=\frac{\Sigma}{2 F^2} \str\, \bigl[ M^\dagger U_0
    \xi(x)^2 + \xi(x)^2 U_0^{-1} M \bigr], \\
  \label{eq:L2C}
  {\cal L}^C_2 &= -\frac 12 \str  U_0^{-1} C U_0
    [\xi(x),[C,\xi(x)]] -\frac i2 \str\, (U_0^{-1} C
    U_0+C) [\xi(x), \partial_0\xi(x)]\,, \\
  {\cal L}^N_2 &= \frac{1}{12F^2}\str\,
    [\partial_\rho\xi(x),\xi(x)][\partial_\rho\xi(x),\xi(x)]
    -\frac1{3\sqrt2F} \str \, U_0^{-1} [C, U_0] [\xi(x),
    [\partial_0\xi(x), \xi(x)]] \,.
\end{align}
In this section we will integrate out the fluctuations in $\xi$ in
order to obtain an effective finite-volume partition function.  The
term ${\cal L}^M_2$ couples to $U_0$ and $M$, and thus corrects the
leading-order mass term.  In section~\ref{sec:fvcsigma} we discuss its
effect on the low-energy constant $\Sigma$.  The term ${\cal L}^C_2$
couples to $U_0$ and $C$ and corrects the leading-order chemical
potential term.  Its effect on the low-energy constant $F$ is
discussed in section~\ref{sec:fvcf}.  The first term in ${\cal L}^N_2$
can be ignored since it does not couple to $U_0$ and therefore only
amounts to an overall factor in the effective finite-volume partition
function.  The second term in ${\cal L}^N_2$ can be ignored at the
order at which we are working since it does not give rise to
leading-order corrections to $\Sigma$ or $F$.

The integration measure for the parametrization of
Eq.~\eqref{eqn:defu} is of the form
\begin{align}
  d[U] = d[U_0] d[\xi] {\cal J}(\xi)\,,
\end{align}
where $d[U_0]$ is the invariant measure for the constant-mode
integral, $d[\xi]$ is the flat path integral measure of the fields
$\xi$, and ${\cal J}(\xi)$ is the Jacobian corresponding to the
change of variables of Eq.~\eqref{eqn:defu}.
Since $\xi$ does not contain constant modes the kinetic term
in Eq.~\eqref{eqn:lagrlo} suppresses large fluctuations in $\xi$,
and thus the integrand vanishes at the integration boundaries 
of the $\pi$- and $\pi'$-fields.
Therefore the invariant integration measure is well-defined
and there are no anomalous contributions by Efetov-Wegner terms
\cite{Rothstein:1987zz,Constantinescu:1989zz}.  The Jacobian
 must be of the form
\begin{align}\label{eqn:defjac}
  {\cal J}(\xi) = 1 + {\cal O}(\varepsilon^2)
\end{align}
since there can be no contribution from a linear term in $\xi$ because
of $\int d^4x\, \xi(x)=0$.  Thus, at next-to-leading order
the Jacobian only contributes an overall factor to the
effective finite-volume partition function.\footnote{
At higher orders in $\varepsilon$ the effects of the Jacobian can
no longer be absorbed in an overall prefactor of the partition function.}

\subsection{The propagator}

The kinetic term of the Lagrangian in terms of the fields $\pi$,
$\pi'$, $\phi$, $\bar\kappa$, and $\kappa$ is given by
\begin{align}\notag
  \frac{1}{2} \str \left[(\partial_\rho\xi) (\partial_\rho\xi)\right] &=
\frac{1}{2} \tr \left[(\partial_\rho \pi)(\partial_\rho \pi)\right]
+\frac{1}{2} \tr \left[(\partial_\rho \pi')(\partial_\rho \pi')\right]
+\frac{1}{2} \tr \left[(\partial_\rho \phi)(\partial_\rho \phi)\right] 
\\ &\quad+  (\partial_\rho \bar \kappa_{ji})(\partial_\rho \kappa_{ji})\,.
\end{align}
Since the mass term ${\cal L}_2^M$ of the Lagrangian, see
\eqref{eq:L2M}, is of order $\O(\varepsilon^2)$, the fields are
effectively massless.  The massless propagator without zero modes,
$\bar\Delta(x)$, is finite in dimensional regularization~\cite{Hasenfratz:1989pk}.
In appendix~\ref{app:prop} we give explicit
expressions for the relevant propagators used in this work.  For the
pion fields $\pi$ and $\pi'$ the propagators are given by
\cite{Hansen:1990un,Akemann:2008vp}
\begin{align}
  \braket{ \pi(x)_{ab}\pi(y)_{cd} }_0 &= \bar\Delta(x-y) \left[
    \delta_{ad}\delta_{bc} - \frac1{N_f+N_v}\delta_{ab} \delta_{cd} \right],\\
  \braket{ \pi'(x)_{ab}\pi'(y)_{cd} }_0 &= \bar\Delta(x-y) \left[
    \delta_{ad}\delta_{bc} - \frac1{N_v}\delta_{ab} \delta_{cd} \right],
\end{align}
where the average is defined by
\begin{align}
  \braket{\O[\xi]}_0 = \frac{\int d[\xi]\: \O[\xi]\: e^{-\int d^4x\,
      \L_0}} {\int d[\xi]\: e^{-\int d^4x\:\L_0}}\,.
\end{align}
For the scalar field $\phi$ and for the fermionic field $\kappa$ the
propagators are easily shown to be
\begin{align}
  \braket{ \bar\kappa(x)_{ab} \kappa(y)_{cd} }_0 &= -\bar\Delta(x-y)
  \delta_{ac} \delta_{bd}\,,\\ 
  \braket{ \phi(x) \phi(y) }_0 &= \bar\Delta(x-y)\,.
\end{align}
Using the identities
\begin{align}
  \frac1{N_f+N_v}+\frac{N_v^2}{(N_f+N_v)N_f N_v} &= \frac1{N_f}\,,\\
  -\frac1{N_v}+\frac{(N_f+N_v)^2}{(N_f+N_v)N_f N_v} &= \frac1{N_f}\,,
\end{align}
we thus find the propagator of the composite field $\xi$ to be
\begin{align}
  \label{eqn:propagator}
  \braket{ \xi(x)_{ab} \xi(y)_{cd} }_0 &= \bar\Delta(x-y) \left[
    \delta_{ad}\delta_{bc} (-1)^{\varepsilon_b} -
    \frac1{N_f}\delta_{ab} \delta_{cd} \right]
\end{align}
with
\begin{align}
  \varepsilon_b =
  \begin{cases}
    0 & \text{for } 1\le b \le N_f + N_v\,,\\
    1 & \text{for } N_f + N_v < b \le N_f + 2 N_v\,.
  \end{cases}
\end{align}
Note that there is no explicit dependence on the number $N_v$ of
valence quarks in this propagator.

\subsection[Finite-volume corrections to $\Sigma$]{\boldmath
  Finite-volume corrections to $\Sigma$}\label{sec:fvcsigma}

We now integrate out the fluctuations in the $\mathcal{O}(\epsilon^2)$
mass term ${\cal L}_2^M$ to obtain the finite-volume corrections to
the leading-order mass term in ${\cal L}_0$.  Using
\eqref{eqn:propagator} it is straightforward to show that
\begin{align}\label{eqn:defconi}
 \braket{ \str [ A \xi(x) B \xi(y) ] }_0 =\bar\Delta(x-y) \left[\str A \str B  - \frac1{N_f} \str AB \right].
\end{align}
By expanding the action we find that the term
\begin{align}
  \int d^4 x \: \braket{\frac{\Sigma}{2 F^2} \str\, \bigl[ M^\dagger U_0
    \xi(x)^2 + \xi(x)^2 U_0^{-1} M \bigr]}_0
\end{align}
corrects the leading-order mass term in the Lagrangian,
\begin{align}
  -\frac\Sigma 2\str\,\bigl[ M^\dagger U_0 + U_0^{-1} M \bigr]\,,
\end{align}
to
\begin{align}
  -\frac\Sigma2\biggl[1 - \frac{N_f^2-1}{N_f F^2} \bar\Delta(0)\biggr]
  \str\, \bigl[ M^\dagger U_0 + U_0^{-1} M \bigr].
\end{align}
Thus at next-to-leading order we can read off an effective
low-energy constant $\Sigma_\text{eff}$ given by
\begin{align}\label{eqn:seff}
  \frac{\Sigma_\text{eff}}{\Sigma} = 1 - \frac{N_f^2-1}{N_f F^2}
  \bDelta(0)\,.
\end{align}
This is the same result as previously derived for the unquenched
partition function \cite{Gasser:1987ah,Damgaard:2007xg,Akemann:2008vp}.  
It can be shown that at next-to-next-to-leading order (NNLO)
it is no longer possible to absorb the effects of the finite volume in an
effective low-energy constant $\Sigma_\text{eff}$.

\subsection[Finite-volume corrections to $F$]{\boldmath Finite-volume
  corrections to $F$}\label{sec:fvcf}

The calculation of the finite-volume corrections to $F$ is slightly
more involved.  The non-vanishing corrections to the leading-order
imaginary chemical potential term are given by Eq.~\eqref{eq:L2C}.  We
first calculate the contribution of the first term in \eqref{eq:L2C},
\begin{align}
  \label{eq:L2C1}
  -\frac 12 \int d^4x & \braket{\str \, U_0^{-1} C U_0
      [\xi(x),[C,\xi(x)]] }_0 \notag\\ 
  &=-\frac 12 \int d^4x \braket{\str \, U_0^{-1} C U_0 [2\xi(x) C
      \xi(x) - \xi(x)^2 C - C \xi(x)^2 ] }_0 \notag\\
  &=- V \bDelta(0) \left[(\str C)^2-N_f \str U_0^{-1} C U_0 C\right],
\end{align}
where we have used \eqref{eqn:defconi}.  The first term in
\eqref{eq:L2C1} couples only to $C^2$ and thus amounts only to a
prefactor in the effective finite-volume partition function.  The
correction to the leading-order Lagrangian obtained from
\eqref{eq:L2C1} is thus given by
\begin{align}
  \label{eq:f1}
  \frac {\bDelta(0)}{ 2 } N_f  \str\, [C,U_0^{-1}][C, U_0]\,.
\end{align}
The contribution of the second term in \eqref{eq:L2C} is given by
\begin{align}
  -\frac i2 \int d^4x \braket{\str\, (U_0^{-1} C U_0+C) [\xi(x),
      \partial_0\xi(x)] }_0 \sim \partial_0\bDelta(0) = 0
\end{align}
due to the symmetry $\bDelta(x) = \bDelta(-x)$.  However, the square
of this term gives a nonzero contribution.  We need to calculate
\begin{align}\notag
  -\frac12 \biggl\langle\Bigl( -\frac i2 & \int d^4x\, \str\,
  (U_0^{-1} C U_0+C) [\xi(x), \partial_0\xi(x)]
  \Bigr)^2\biggr\rangle_0 \\ 
  &=\frac18 \int d^4x \int d^4y\: \bigl\langle\str\bigl( Y [\xi(x),
    \partial_0\xi(x)] \bigr) \str\bigl( Y [\xi(y),
    \partial_0\xi(y)] \bigr)\bigr\rangle_0
  \label{eq:L2C2}
\end{align}
with $Y = U_0^{-1} C U_0+C$.  After performing all relevant
contractions using \eqref{eqn:propagator} we find
\begin{align}
  \bigl\langle \str [ Y \xi(x) & \xi(x') ]\str [ Y \xi(y) \xi(y') ]
  \bigr\rangle_0 \notag\\ 
  &= \bDelta(x-x')\bDelta(y-y')\Bigl[ (\str Y)^2 N_f^2 -2(\str Y)^2
  +\frac1{N_f^2}(\str Y)^2\Bigr] \notag\\
  &\quad+ \bDelta(x-y)\bDelta(x'-y') \Bigl[(\str Y)^2 - \frac 2{N_f}\str
  Y^2 +\frac1{N_f^2} (\str Y)^2 \Bigr] \notag\\
  &\quad+ \bDelta(x-y')\bDelta(x'-y) \Bigl[N_f \str Y^2 - \frac
  2{N_f}\str Y^2 +\frac1{N_f^2} (\str Y)^2 \Bigr]\,. 
  \label{eq:stryy}
\end{align}
Since $\str Y=2\str C$ does not couple to $U_0$ we only need to take
into account the terms involving $\str Y^2$.  We denote the irrelevant
terms by ``$\ldots$'' and write
\begin{align}
  \eqref{eq:stryy}=-\frac{\str Y^2}{N_f}[2\bDelta(x-y)\bDelta(x'-y')
  +(2-N_f^2)\bDelta(x-y')\bDelta(x'-y)] + \ldots
\end{align}
We need to calculate
\begin{align}
  (\partial_{x'_0} & - \partial_{x_0})(\partial_{y'_0} - \partial_{y_0})
  \braket{ \str [ Y \xi(x) \xi(x') ] \str [ Y \xi(y) \xi(y')
    ]}_0\Big\vert_{x=x',\ y=y'}  \notag\\
  &=-2 N_f\str Y^2 \left[\bigl(\partial_{0}\bDelta(x-y)\bigr)
    \bigl(\partial_{0}\bDelta(x-y)\bigr)-\bigl(\partial_{0}^2\bDelta(x-y)\bigr)
    \bDelta(x-y)\right] + \ldots
\end{align}
Thus we find
\begin{align}
  \eqref{eq:L2C2}=-\frac V2 N_f \str Y^2 \int d^4x \:
  \bigl(\partial_0\bDelta(x)\bigr)^2 + \ldots\,,
\end{align}
where we have used the fact that the propagator is periodic in time.
Therefore the corrections to the effective Lagrangian are given by
\begin{align}
  \label{eq:f2}
  - \frac12 N_f\str [C,U_0^{-1}][C, U_0] \int d^4x \:
  \bigl(\partial_0\bDelta(x)\bigr)^2\,.
\end{align}
Combining \eqref{eq:f1} and \eqref{eq:f2}, we find that the
fluctuations correct the leading-order contribution to the Lagrangian,
\begin{align}
  -\frac{F^2}4\str\, [C, U_0^{-1}] [C, U_0]\,,
\end{align}
to
\begin{align}
  -\frac{F^2}4\str\,[C, U_0^{-1}] [C, U_0]
  \left[1-\frac {2 N_f}{F^2}\Bigl(\bDelta(0)
      -\int d^4x \bigl(\partial_0\bDelta(x)\bigr)^2\Bigr) \right].
\end{align}
Thus at next-to-leading order we find an effective low-energy constant
$F_\text{eff}$ given by
\begin{align}\label{eqn:feff}
  \frac{F_\text{eff}}{F} = 1 -\frac {N_f}{F^2}\Bigl(\bDelta(0)-\int
    d^4x \: \bigl(\partial_0\bDelta(x)\bigr)^2\Bigr)\,.
\end{align}
This again agrees with the result for the unquenched partition
function \cite{Damgaard:2007xg,Akemann:2008vp}.  As in the case of $\Sigma$,
at NNLO it is no longer possible to absorb the effects of the finite volume
in an effective low-energy constant $F_\text{eff}$.

\subsection{The universal limit}\label{sec:universal}

In this section we concern ourselves with the limit $V \to \infty$
while keeping $MV\Sigma \sim \mathcal{O}(\epsilon^0)$.  It is well
known that QCD in this limit behaves in a universal way and agrees
with chiral RMT.  It was first shown in
Refs.~\cite{Osborn:1998qb,Damgaard:1998xy} how universal results for
the Dirac spectrum (obtained earlier in RMT) can be derived from the
effective low-energy theory.  In the following we show that the
correct universal limit also follows from the effective theory in the
formulation of Sharpe and Shoresh described above.  For simplicity we
restrict ourselves to the case of vanishing imaginary chemical
potential, $C=0$.

Since the fluctuations in $\xi$ are suppressed for $V \to \infty$ only
the zero-mode integral survives in this limit, and the partition
function for fixed topological charge $\nu$ is given by
\begin{align}
  Z_\nu &= \int d[U_0]\: \sdet^\nu(U_0) \exp\,\biggl(
  \frac{\Sigma V}{2} \str\,\bigl[ M^\dagger U_0 + U_0^{-1}M\bigr]\biggr)\,,
\end{align}
where the integration manifold is specified in Eq.~\eqref{eqn:defgaf}.
There are different methods to calculate integrals over
supermanifolds, see, e.g.,
\cite{Guhr:1996vx,Guhr:1996mx,Lehner:2008fp}.  In our case it is
sufficient to choose an explicit parametrization and reduce the
integral to ordinary group integrals.  For convenience we use a
slightly different notation and calculate
\begin{align}
  Z_\nu &= \int d[U]\: \sdet^\nu(U) \exp\,\Bigl(
  \str\,\bigl[M^\dagger U + U^{-1} M\bigr]\Bigr)
\end{align}
with integration manifold given by
\begin{align}
  U =
  \begin{pmatrix}
    V e^{N_v \phi} & 0 \\ 0 & V' e^{(N_f+N_v) \phi}
  \end{pmatrix}
  \exp
  \begin{pmatrix}
    0 & \bar\kappa^T \\
    \kappa & 0
  \end{pmatrix}
  \equiv U_c U_g\,,
\end{align}
where $V \in \U(N_f+N_v)$, $V' \in \Gl(N_v)/\U(N_v)$ with $\det V' =
1$, and $\phi \in \R$.  Thus we have
\begin{align}
  \sdet U = \det V \equiv e^{i \theta}
\end{align}
with $\theta \in [0,2\pi)$.  This is the zero-mode integral following
from the parametrization used in the perturbative calculation above.
In the literature a similar integral was computed to determine the
static limit of partially quenched chiral perturbation theory
\cite{Osborn:1998qb,Damgaard:1998xy} that amounts to replacing $U_c$
by
\begin{align}
  U_c \to \begin{pmatrix}
    V & 0 \\ 0 & V' e^{\phi/N_v}
  \end{pmatrix}.
\end{align}
Note first that a parametrization such as $U=U_c U_g$ above leads to
factorization of the corresponding measure as
\begin{align}
 d[U]=d[U_c] d[U_g]\,.
\end{align}
This is due to the fact that the invariant length element is
\begin{align}
  ds^2 &= \str\, [ dU d(U^{-1}) ] \notag\\
  &= \str\, [dU_c d(U_c^{-1}) + dU_g d(U_g^{-1}) 
-2 U_c^{-1} dU_c dU_g U_g^{-1}] \notag\\
  &=\str\, [dU_c d(U_c^{-1}) + dU_g d(U_g^{-1})]
\end{align}
since
\begin{align}
  dU_g U_g^{-1} =
  \begin{pmatrix}
    0 & d\bar\kappa^T \\
    d\kappa & 0
  \end{pmatrix},
\end{align}
$U_c^{-1}dU_c $ is block diagonal, and therefore
\begin{align}
  \str [U_c^{-1} dU_c dU_g U_g^{-1}] = 0\,.
\end{align}
In both parametrizations the measure of $V$, $V'$, and $\phi$ also
factorizes.  Thus 
\begin{align}
  d[U]=d[U_g]d[V]d[V']d\phi
\end{align}
in both cases.  Note that this parametrization has no contributions
from Efetov-Wegner terms, as was discussed in a special case in the
literature \cite{Damgaard:1998xy}.  Introducing the short-hand
notation 
\begin{align}
  U_g M^\dagger=
  \begin{pmatrix}
    X_\text{ff} & X_\text{fb} \\
    X_\text{bf} & X_\text{bb}
  \end{pmatrix},\qquad
  M U_g^{-1}=
  \begin{pmatrix}
    Y_\text{ff} & Y_\text{fb} \\
    Y_\text{bf} & Y_\text{bb}
  \end{pmatrix},
\end{align}
we find for the first parametrization
\begin{align}
  \str\,&[M^\dagger U + M U^{-1}] 
  = \str\,[M^\dagger U_c U_g + M U_g^{-1} U_c^{-1}] 
  = \str\,[U_c X + U_c^{-1} Y] \notag \\
  & = \tr\,\bigl[V e^{N_v \phi} X_\text{ff} - V' e^{(N_f+N_v) \phi}
  X_\text{bb} + V^{-1} e^{-N_v \phi} Y_\text{ff} - V'^{-1}
  e^{-(N_f+N_v) \phi} Y_\text{bb} \bigr]\,.
\end{align}
Next we use a result of \cite{Schlittgen:2002tj},
\begin{align}
  \int_{\U(p)}\!\! d[U]\: {\det}^\nu(U) \exp\left[\tr(AU+BU^{-1}) \right]\!
  = c_p \det(BA^{-1})^{\nu/2} \: \frac{\det\bigl[\mu_i^{j-1}
    I_{\nu+j-1}(2\mu_i)\bigr]}{\Delta(\mu^2)}\,, 
\end{align}
where $c_p$ is a constant, $\Delta(\mu^2)$ is the Vandermonde
determinant, and the $\mu_i^2$ are the eigenvalues of $AB$.  Thus the
integral over $V$ results in
\begin{align}
  e^{-N_v(N_f+N_v)\nu \phi} \det(Y_\text{ff}X_\text{ff}^{-1})^{\nu/2}
  \:\frac{\det\bigl[\mu_i^{j-1} I_{\nu+j-1}(2\mu_i)\bigr]}{\Delta(\mu^2)} 
\end{align}
with $\mu_i^2$ the eigenvalues of $X_\text{ff}Y_\text{ff}$.  In the
second parametrization we find
\begin{align}
  \str\,[M^\dagger U + M U^{-1}] &=\tr\,\bigl[
  V X_\text{ff} - V' e^{\phi/N_v} X_\text{bb}
  +V^{-1} Y_\text{ff} - V'^{-1} e^{-\phi/N_v} Y_\text{bb} \bigr]\,.
\end{align}
Note that in this parametrization we also have an additional factor of
$e^{-\phi\nu}$ from the superdeterminant.  Thus the integral over $V$
leads to
\begin{align}
  e^{-\nu \phi} \det(Y_\text{ff}X_\text{ff}^{-1})^{\nu/2}\:
  \frac{\det\bigl[\mu_i^{j-1} I_{\nu+j-1}(2\mu_i)\bigr]}{\Delta(\mu^2)} 
\end{align}
with $\mu_i^2$ already defined above.  Now we let $\phi \to \phi
N_v(N_f+N_v)$ in order to have the same prefactor of $V'$ and
$V'^{-1}$ in the supertrace.
In both parametrizations the resulting integral is
\begin{align}
  \int d[U_g] d[V'] \int_{-\infty}^\infty d\phi \: & e^{-\nu \phi (N_f+N_v)N_v} 
  \det(Y_\text{ff}X_\text{ff}^{-1})^{\nu/2} \:
  \frac{\det\bigl(\mu_i^{j-1} I_{\nu+j-1}(2\mu_i)\bigr)}{\Delta(\mu^2)} 
  \notag\\
  &\times \exp\left(-\tr\left[V' e^{(N_f+N_v)\phi} X_\text{bb}+
      V'^{-1} e^{-(N_f+N_v)\phi} Y_\text{bb}\right]\right). 
\end{align}
This completes the matching with
Refs.~\cite{Osborn:1998qb,Damgaard:1998xy} and is sufficient to show
that the parametrization of the NG manifold used in this work leads to
the correct universal limit.

In order to extend this proof to the general case of $C\ne 0$
we would need to calculate the group integral
\begin{align}\label{eqn:newgi}
  \int_{\U(p)} d[U]\: {\det}^\nu(U) \exp\left[\tr(AU+BU^{-1})+\tr(D U D U^{-1}) \right],
\end{align}
where $A$, $B$, and $D$ are arbitrary complex $p \times p$-matrices.
This, however, is beyond the scope of this work.

\section{Conclusions}\label{sec:concl}

In this work we have calculated the partially quenched partition
function of QCD at next-to-leading order in the
$\varepsilon$-expansion at nonzero imaginary chemical potential.  
We considered a theory with $N_f+N_v$ fermionic quarks and
$N_v$ bosonic quarks, as formulated by Sharpe and Shoresh
\cite{Sharpe:2001fh}, in a finite volume $V$ with microscopic
quark masses $M$, i.e., $M V \Sigma = {\cal O}(\varepsilon^0)$.  The
knowledge of the analytic form of the partially quenched partition
function suffices to obtain all spectral correlation functions of the
Dirac operator $\Ds$.  In this sense our results for the
finite-volume behavior of the theory hold universally for all
observables that can be obtained from spectral correlation functions
of $\Ds$.  We found that the partially quenched partition function 
has the same finite-volume corrections as the unquenched
partition function of QCD with $N_f$ quarks, i.e., at next-to-leading
order in $\varepsilon$ there are effective low-energy constants
$\Sigma_\text{eff}$ and $F_\text{eff}$,
\begin{align}\tag{\ref{eqn:seff}}
  \frac{\Sigma_\text{eff}}{\Sigma} &= 1 - \frac{N_f^2-1}{N_f F^2}
  \bDelta(0)\,,\\ \tag{\ref{eqn:feff}}
  \frac{F_\text{eff}}{F} &= 1 -\frac {N_f}{F^2}\Bigl(\bDelta(0)-\int
    d^4x \: \left(\partial_0\bDelta(x)\right)^2\Bigr)\,,
\end{align}
where $\bDelta(x)$ is the massless propagator.  In
appendix~\ref{app:prop} we give closed formulas for the relevant
propagators in dimensional regularization and numerical values for
typical geometries.  As a side result of our calculation we showed
that the constant-mode integral of this theory agrees with previous
results from random matrix theory.  Therefore the correct universal
limit is obtained at $V \to \infty$.  Note  that the proof
was only given for vanishing chemical potential and that the knowledge
of the group integral \eqref{eqn:newgi} is needed to complete the
proof also for nonzero chemical potential.

\FIGURE[t]{
  \centering
  \includegraphics[width=7.4cm]{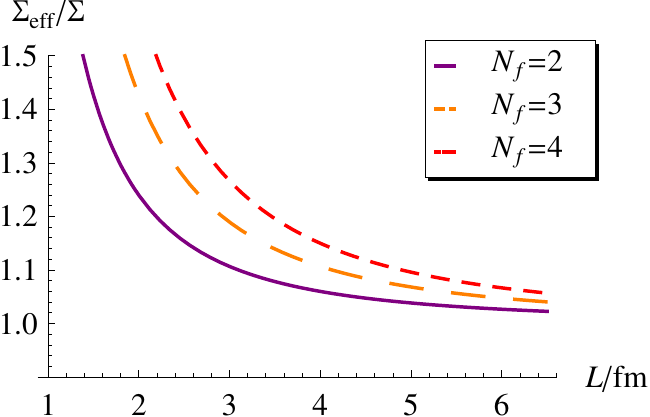}
  \includegraphics[width=7.4cm]{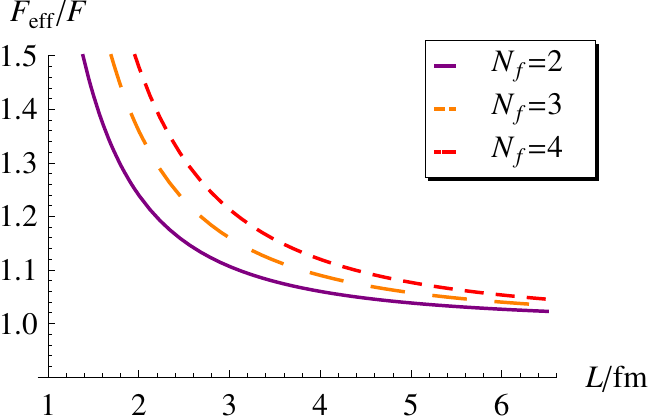}
  \caption{Volume-dependence at NLO of the low-energy constants $\Sigma_\text{eff}$ (left)
    and $F_\text{eff}$ (right) in a symmetric box with dimensions $L_0=L_1=L_2=L_3=L$ at $F=90$ MeV.}
  \label{fig:p}
}
In figure~\ref{fig:p} we show the finite-volume corrections at NLO
to the low-energy constants $\Sigma$ and $F$ as a function of the box size $L$
in a symmetric box.  Note that the effects of the finite volume increase with the number
of sea quark flavors $N_f$ and that, depending on $N_f$,
a box size of $3 - 5$ fm is necessary to reduce the effects of the finite volume at NLO to about 10\%.
The effects are calculated at $F=90$ MeV.
\FIGURE[t]{
  \centering
  \includegraphics[width=7.3cm]{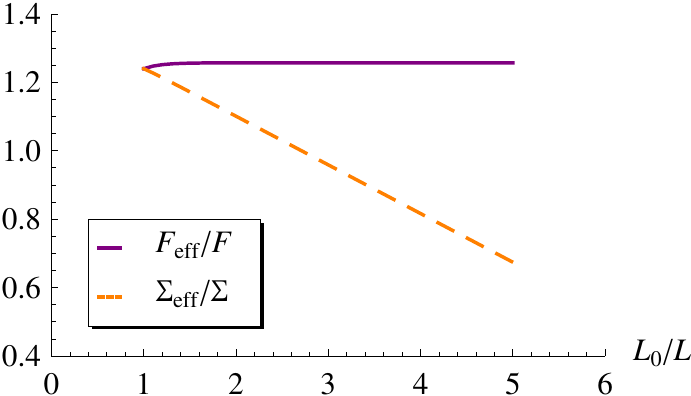}
  \hspace{0.2cm}
  \includegraphics[width=7.3cm]{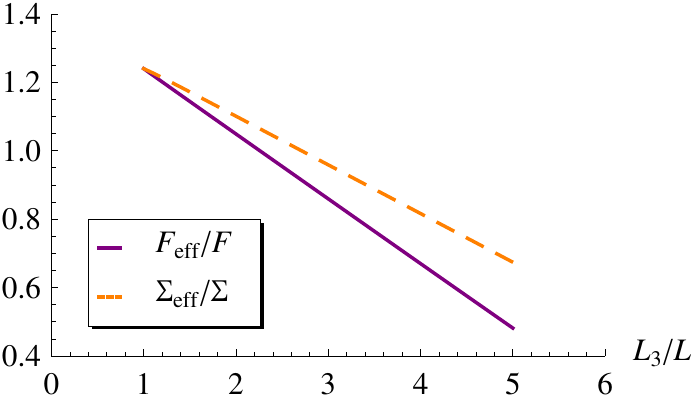}
  \caption{Effect of an asymmetric box with parameters $N_f=2$, $L=2$ fm, and $F=90$ MeV.
    We compare a large temporal dimension $L_0$ with $L_1=L_2=L_3=L$ (left) to
    a large spatial dimension $L_3$ with $L_0=L_1=L_2=L$ (right).
  }
  \label{fig:pratio}
}
In figure~\ref{fig:pratio} we show the effect of an asymmetric box with $N_f=2$ and $L=2$ fm.
An important message of this figure is that the magnitude of the finite-volume corrections can be
significantly reduced by choosing one large spatial dimension instead of a large temporal dimension.  
The reason for this behavior is that the chemical potential only affects the temporal direction,
see Eq.~\eqref{eqn:nablachem}, and therefore breaks the permutation symmetry of the
four dimensions.  This manifests itself in the propagator
\begin{align}
  \int d^4x \left(\partial_0 \bar\Delta(x)\right)^2
\end{align}
which, as shown in Eq.~\eqref{eqn:knn},
contains a term proportional to $L_0^2 / \sqrt V$, where $L_0$ is the size of the temporal
dimension.  This term leads to an enhancement of the corrections in case of a large temporal
dimension.
Choosing instead one large spatial dimension, the finite-volume corrections are reduced,
unless the asymmetry is too large.  For the parameters used in
figure~\ref{fig:pratio}, the optimal value is $L_3/L \approx 2$.

This is good news.  Many lattice simulations (at zero chemical
potential) are performed with $L_1=L_2=L_3=L$ and $L_0=2L$.  To
determine $F$, it suffices to introduce the imaginary chemical
potential in the valence sector.  Therefore, one can take a suitable
set of existing dynamical configurations and redefine
$L_0\leftrightarrow L_3$ before adding the chemical potential.\footnote{
Note that this procedure increases the temperature of the
system by a factor of two.  One needs to check that the system does not end up
in the chirally restored phase, in which our results no longer apply.}  This
will minimize the finite-volume corrections for both $\Sigma$ and $F$,
at least for the parameter values chosen in figure~\ref{fig:pratio}.

\acknowledgments 

We thank Hidenori Fukaya and Shoji Hashimoto for stimulating
discussions and the Theory Group of the INPS at KEK Tsukuba for their
hospitality.  This work was supported in part by BayEFG (CL) and by
DFG and KEK (TW).

\appendix

\section{The massless propagator in dimensional
  regularization}\label{app:prop}

For convenience we collect in this appendix explicit formulas for the
massless propagator in dimensional regularization, $\bar\Delta(x)$,
that were derived in Refs.~\cite{Hasenfratz:1989pk,Hansen:1990un}.
The two relevant quantities for the finite-volume corrections to
$\Sigma$ and $F$ are given by
\begin{align}
  \bar\Delta(0) = -\frac{\beta_1}{\sqrt V}
\end{align}
and
\begin{align}\label{eqn:knn}
  \int d^4x \left(\partial_0 \bar\Delta(x)\right)^2 = -\frac{1}{2\sqrt
    V}\left[\beta_1-\frac{L_0^2}{\sqrt V} k_{00} \right]\!,
\end{align}
where $\beta_1$ and $k_{00}$ are so-called shape coefficients, i.e.,
they only depend on the quantities $l_i = L_i / V^{1/4}$ with $i=0,1,2,3$.
The shape coefficient $\beta_1$ is given by
\begin{align}
  \beta_1 &= \frac{1}{4\pi} \left[2-\hat\alpha_{-1}(l_j) -
    \hat\alpha_{-1}(l_j^{-1})\right]
\end{align}
with
\begin{align}
  \hat \alpha_{-1} (x_j) =
  \int_0^1 dt \: t^{-2}\Biggl[ \prod_{j=0}^3 S(x_j^2/t) - 1 \Biggr],
\end{align}
where $S(x)$ is an elliptic theta-function defined by
\begin{align}
   S(x) &= \sum_{n=-\infty}^\infty e^{-\pi n^2 x}\,.
\end{align}
The shape coefficient $k_{00}$ is given by
\begin{align}
  k_{00} &= \frac1{12}-\sum_{\vec n}\frac{1}{4\sinh(l_0 q_n/2)^2}\,,
\end{align}
where the sum is over all integers $(n_1,n_2,n_3)\ne (0,0,0)$ and
\begin{align}
  q_n^2 =\sum_{j=1}^3 (2\pi n_j/l_j)^2\,.
\end{align}
\TABLE[t]{
  \centering
  \begin{tabular}[t]{l|cccc}
    $L_0 / L$ & $1$ & $2$ & $3$ & $4$ \\
    \hline
    $\beta_1$ & $0.1404610$ & $0.0836011$ & $-0.0419417$ & $-0.215097$ \\
    $k_{00}$ & $0.0702305$ & $0.0833122$ & $0.0833333$ & $0.0833333$
  \end{tabular}
  \caption{Coefficients for an asymmetric box with $L_1=L_2=L_3=L$ and temporal dimension $L_0$.}
  \label{tab:sct}
}
\TABLE[t]{
  \centering
  \begin{tabular}[t]{l|cccc}
    $L_3 / L$ & $1$ & $2$ & $3$ & $4$ \\
    \hline
    $\beta_1$ & $0.1404610$ & $0.0836011$ & $-0.0419417$ & $-0.215097$ \\
    $k_{00}$ & $0.0702305$ & $-0.0322630$ & $-0.2984300$ & $-0.731240$
  \end{tabular}
  \caption{Coefficients for an asymmetric box with $L_0=L_1=L_2=L$ and spatial dimension $L_3$.
    Note that $\beta_1$ is symmetric under the exchange of the temporal with a spatial dimension.}
  \label{tab:scs}
}
In tables~\ref{tab:sct} and \ref{tab:scs} we give numerical values for common
shapes.

\bibliographystyle{JHEP}
\bibliography{pqchpt} 

\providecommand{\href}[2]{#2}\begingroup\raggedright\begin{thebibliography}{10}

\bibitem{Gasser:1987ah}
J.~Gasser and H.~Leutwyler, {\it Thermodynamics of chiral symmetry},  {\em
  Phys. Lett.} {\bf B188} (1987) 477.

\bibitem{Leutwyler:1992yt}
H.~Leutwyler and A.~V. Smilga, {\it Spectrum of {Dirac} operator and role of
  winding number in {QCD}},  {\em Phys. Rev.} {\bf D46} (1992) 5607--5632.

\bibitem{Shuryak:1992pi}
E.~V. Shuryak and J.~J.~M. Verbaarschot, {\it Random matrix theory and spectral
  sum rules for the {Dirac} operator in {QCD}},  {\em Nucl. Phys.} {\bf A560}
  (1993) 306--320, [\href{http://xxx.lanl.gov/abs/hep-th/9212088}{{\tt
  hep-th/9212088}}].

\bibitem{Verbaarschot:2000dy}
J.~J.~M. Verbaarschot and T.~Wettig, {\it Random matrix theory and chiral
  symmetry in {QCD}},  {\em Ann. Rev. Nucl. Part. Sci.} {\bf 50} (2000)
  343--410, [\href{http://xxx.lanl.gov/abs/hep-ph/0003017}{{\tt
  hep-ph/0003017}}].

\bibitem{Akemann:2007rf}
G.~Akemann, {\it Matrix models and {QCD} with chemical potential},  {\em Int.
  J. Mod. Phys.} {\bf A22} (2007) 1077--1122,
  [\href{http://xxx.lanl.gov/abs/hep-th/0701175}{{\tt hep-th/0701175}}].

\bibitem{Damgaard:2005ys}
P.~H. Damgaard, U.~M. Heller, K.~Splittorff, and B.~Svetitsky, {\it A new
  method for determining {F}(pi) on the lattice},  {\em Phys. Rev.} {\bf D72}
  (2005) 091501, [\href{http://xxx.lanl.gov/abs/hep-lat/0508029}{{\tt
  hep-lat/0508029}}].

\bibitem{Akemann:2006ru}
G.~Akemann, P.~H. Damgaard, J.~C. Osborn, and K.~Splittorff, {\it A new chiral
  two-matrix theory for {Dirac} spectra with imaginary chemical potential},
  {\em Nucl. Phys.} {\bf B766} (2007) 34--67,
  [\href{http://xxx.lanl.gov/abs/hep-th/0609059}{{\tt hep-th/0609059}}].

\bibitem{Damgaard:2007xg}
P.~H. Damgaard, T.~DeGrand, and H.~Fukaya, {\it Finite-volume correction to the
  pion decay constant in the epsilon-regime},  {\em JHEP} {\bf 12} (2007) 060,
  [\href{http://xxx.lanl.gov/abs/0711.0167}{{\tt arXiv:0711.0167}}].

\bibitem{Akemann:2008vp}
G.~Akemann, F.~Basile, and L.~Lellouch, {\it Finite size scaling of meson
  propagators with isospin chemical potential},  {\em JHEP} {\bf 12} (2008)
  069, [\href{http://xxx.lanl.gov/abs/0804.3809}{{\tt arXiv:0804.3809}}].

\bibitem{Edwards:1975zz}
S.~F. Edwards and P.~W. Anderson, {\it Theory of spin glasses},  {\em J. Phys.}
  {\bf F5} (1975) 965--974.

\bibitem{Kamenev:1999aa}
A.~Kamenev and M.~Mezard, {\it {Wigner}-{Dyson} statistics from the replica
  method},  {\em J. Phys. A} {\bf 32} (1999) 4373--4388,
  [\href{http://xxx.lanl.gov/abs/cond-mat/9901110}{{\tt cond-mat/9901110}}].

\bibitem{Kanzieper:2002ix}
E.~Kanzieper, {\it Replica field theories, {Painleve} transcendents and exact
  correlation functions},  {\em Phys. Rev. Lett.} {\bf 89} (2002) 250201,
  [\href{http://xxx.lanl.gov/abs/cond-mat/0207745}{{\tt cond-mat/0207745}}].

\bibitem{Splittorff:2002eb}
K.~Splittorff and J.~J.~M. Verbaarschot, {\it Replica limit of the {Toda}
  lattice equation},  {\em Phys. Rev. Lett.} {\bf 90} (2003) 041601,
  [\href{http://xxx.lanl.gov/abs/cond-mat/0209594}{{\tt cond-mat/0209594}}].

\bibitem{Damgaard:2007ep}
P.~H. Damgaard and H.~Fukaya, {\it Partially quenched chiral perturbation
  theory in the epsilon-regime},  {\em Nucl. Phys.} {\bf B793} (2008) 160--191,
  [\href{http://xxx.lanl.gov/abs/0707.3740}{{\tt arXiv:0707.3740}}].

\bibitem{Bernardoni:2007hi}
F.~Bernardoni and P.~Hernandez, {\it Finite-size scaling for the left-current
  correlator with non-degenerate quark masses},  {\em JHEP} {\bf 10} (2007)
  033, [\href{http://xxx.lanl.gov/abs/0707.3887}{{\tt arXiv:0707.3887}}].

\bibitem{Bernardoni:2008ei}
F.~Bernardoni, P.~H. Damgaard, H.~Fukaya, and P.~Hernandez, {\it Finite volume
  scaling of pseudo {Nambu-Goldstone} bosons in {QCD}},  {\em JHEP} {\bf 10}
  (2008) 008, [\href{http://xxx.lanl.gov/abs/0808.1986}{{\tt
  arXiv:0808.1986}}].

\bibitem{Damgaard:2008zs}
P.~H. Damgaard and H.~Fukaya, {\it The chiral condensate in a finite volume},
  {\em JHEP} {\bf 01} (2009) 052,
  [\href{http://xxx.lanl.gov/abs/0812.2797}{{\tt arXiv:0812.2797}}].

\bibitem{Efetov:1983zz}
K.~B. Efetov, {\it Supersymmetry and theory of disordered metals},  {\em
  Advances in Physics} {\bf 32} (1983) 53--127.

\bibitem{Morel:1987xk}
A.~Morel, {\it Chiral logarithms in quenched {QCD}},  {\em J. Phys. (France)}
  {\bf 48} (1987) 1111--1119.

\bibitem{Bernard:1993sv}
C.~W. Bernard and M.~F.~L. Golterman, {\it Partially quenched gauge theories
  and an application to staggered fermions},  {\em Phys. Rev.} {\bf D49} (1994)
  486--494, [\href{http://xxx.lanl.gov/abs/hep-lat/9306005}{{\tt
  hep-lat/9306005}}].

\bibitem{Sharpe:2001fh}
S.~R. Sharpe and N.~Shoresh, {\it Partially quenched chiral perturbation theory
  without $\phi_0$},  {\em Phys. Rev.} {\bf D64} (2001) 114510,
  [\href{http://xxx.lanl.gov/abs/hep-lat/0108003}{{\tt hep-lat/0108003}}].

\bibitem{DeGrand:2007tm}
T.~DeGrand and S.~Schaefer, {\it Parameters of the lowest order chiral
  {Lagrangian} from fermion eigenvalues},  {\em Phys. Rev.} {\bf D76} (2007)
  094509, [\href{http://xxx.lanl.gov/abs/0708.1731}{{\tt arXiv:0708.1731}}].

\bibitem{Osborn:1998qb}
J.~C. Osborn, D.~Toublan, and J.~J.~M. Verbaarschot, {\it From chiral random
  matrix theory to chiral perturbation theory},  {\em Nucl. Phys.} {\bf B540}
  (1999) 317--344, [\href{http://xxx.lanl.gov/abs/hep-th/9806110}{{\tt
  hep-th/9806110}}].

\bibitem{Damgaard:1998xy}
P.~H. Damgaard, J.~C. Osborn, D.~Toublan, and J.~J.~M. Verbaarschot, {\it The
  microscopic spectral density of the {QCD} {Dirac} operator},  {\em Nucl.
  Phys.} {\bf B547} (1999) 305--328,
  [\href{http://xxx.lanl.gov/abs/hep-th/9811212}{{\tt hep-th/9811212}}].

\bibitem{Fukaya:2007pn}
H.~Fukaya {\em et.~al.}, {\it Lattice study of meson correlators in the
  epsilon-regime of two-flavor {QCD}},  {\em Phys. Rev.} {\bf D77} (2008)
  074503, [\href{http://xxx.lanl.gov/abs/0711.4965}{{\tt arXiv:0711.4965}}].

\bibitem{Bernard:1974bq}
C.~W. Bernard, {\it Feynman rules for gauge theories at finite temperature},
  {\em Phys. Rev.} {\bf D9} (1974) 3312.

\bibitem{Akemann:2003tv}
G.~Akemann and P.~H. Damgaard, {\it Distributions of {Dirac} operator
  eigenvalues},  {\em Phys. Lett.} {\bf B583} (2004) 199--206,
  [\href{http://xxx.lanl.gov/abs/hep-th/0311171}{{\tt hep-th/0311171}}].

\bibitem{Zirnbauer:1996zz}
M.~R. Zirnbauer, {\it Riemannian symmetric superspaces and their origin in
  random-matrix theory},  {\em J. Math. Phys.} {\bf 37} (1996) 4986--5018.

\bibitem{Dalmazi:2000bs}
D.~Dalmazi and J.~J.~M. Verbaarschot, {\it The replica limit of unitary matrix
  integrals},  {\em Nucl. Phys.} {\bf B592} (2001) 419--444,
  [\href{http://xxx.lanl.gov/abs/hep-th/0005229}{{\tt hep-th/0005229}}].

\bibitem{Rothstein:1987zz}
M.~J. Rothstein, {\it Integration on noncompact supermanifolds},  {\em
  Transactions of the American Mathematical Society} {\bf 299} (1987) 387--396.

\bibitem{Constantinescu:1989zz}
F.~Constantinescu and H.~F. de~Groote, {\it The integral theorem for
  supersymmetric invariants},  {\em J. Math. Phys.} {\bf 30} (1989) 981--992.

\bibitem{Hasenfratz:1989pk}
P.~Hasenfratz and H.~Leutwyler, {\it Goldstone boson related finite size
  effects in field theory and critical phenomena with {O(N)} symmetry},  {\em
  Nucl. Phys.} {\bf B343} (1990) 241--284.

\bibitem{Hansen:1990un}
F.~C. Hansen, {\it Finite size effects in spontaneously broken {SU(N)} x
  {SU(N)} theories},  {\em Nucl. Phys.} {\bf B345} (1990) 685--708.

\bibitem{Guhr:1996vx}
T.~Guhr and T.~Wettig, {\it An {Itzykson-Zuber}-like integral and diffusion for
  complex ordinary and supermatrices},  {\em J. Math. Phys.} {\bf 37} (1996)
  6395--6413, [\href{http://xxx.lanl.gov/abs/hep-th/9605110}{{\tt
  hep-th/9605110}}].

\bibitem{Guhr:1996mx}
T.~Guhr, {\it {Gelfand-Tzetlin} coordinates for the unitary supergroup},  {\em
  Comm. Math. Phys.} {\bf 176} (1996) 555--576.

\bibitem{Lehner:2008fp}
C.~Lehner, T.~Wettig, T.~Guhr, and Y.~Wei, {\it Character expansion method for
  supergroups and extended superversions of the {Leutwyler-Smilga} and
  {Berezin-Karpelevich} integrals},  {\em J. Math. Phys.} {\bf 49} (2008)
  063510, [\href{http://xxx.lanl.gov/abs/0801.1226}{{\tt arXiv:0801.1226}}].

\bibitem{Schlittgen:2002tj}
B.~Schlittgen and T.~Wettig, {\it Generalizations of some integrals over the
  unitary group},  {\em J. Phys.} {\bf A36} (2003) 3195--3202,
  [\href{http://xxx.lanl.gov/abs/math-ph/0209030}{{\tt math-ph/0209030}}].

\end{thebibliography}\endgroup

\end{document}